\documentclass[twocolumn,aps,showpacs,preprintnumbers,amsmath,amssymb, superscriptaddress]{revtex4-1}
\pdfoutput=1
\usepackage{bm}
\usepackage{graphicx}
\usepackage{color}
\begin{document}

\title{ARPES observation of isotropic superconducting gaps in isovalent Ru-substituted Ba(Fe$_{0.75}$Ru$_{0.25}$)$_2$As$_2$}

\author{N. Xu}
\affiliation{Beijing National Laboratory for Condensed Matter Physics, and Institute of Physics, Chinese Academy of Sciences, Beijing 100190, China}
\author{P. Richard}\email{p.richard@iphy.ac.cn}
\affiliation{Beijing National Laboratory for Condensed Matter Physics, and Institute of Physics, Chinese Academy of Sciences, Beijing 100190, China}
\author{X. -P. Wang}
\affiliation{Beijing National Laboratory for Condensed Matter Physics, and Institute of Physics, Chinese Academy of Sciences, Beijing 100190, China}
\author{X. Shi}
\affiliation{Beijing National Laboratory for Condensed Matter Physics, and Institute of Physics, Chinese Academy of Sciences, Beijing 100190, China}
\author{A. van Roekeghem}
\affiliation{Beijing National Laboratory for Condensed Matter Physics, and Institute of Physics, Chinese Academy of Sciences, Beijing 100190, China}
\affiliation{Centre de Physique Th\'{e}Žorique, Ecole Polytechnique, CNRS-UMR7644, 91128 Palaiseau, France}
\author{T. Qian}
\affiliation{Beijing National Laboratory for Condensed Matter Physics, and Institute of Physics, Chinese Academy of Sciences, Beijing 100190, China}
\author{E. Ieki}
\affiliation{Department of Physics, Tohoku University, Sendai 980-8578, Japan}
\author{K. Nakayama}
\affiliation{Department of Physics, Tohoku University, Sendai 980-8578, Japan}
\author{T. Sato}
\affiliation{Department of Physics, Tohoku University, Sendai 980-8578, Japan}
\affiliation{TRiP, Japan Science and Technology Agency (JST), Kawaguchi 332-0012, Japan}
\author{E. Rienks}
\affiliation{Helmholtz-Zentrum Berlin, BESSY, D-12489 Berlin, Germany}
\author{S. Thirupathaiah}
\affiliation{Institute for Solid State Research, IFW Dresden, D-01171 Dresden, Germany}
\author{J. Xing}
\affiliation{Center for Superconducting Physics and Materials, National Laboratory for Solid State Microstructures, Department of Physics, Nanjing University, Nanjing 210093, China}
\author{H.-H. Wen}
\affiliation{Center for Superconducting Physics and Materials, National Laboratory for Solid State Microstructures, Department of Physics, Nanjing University, Nanjing 210093, China}
\author{M. Shi}
\affiliation{Paul Scherrer Institut, Swiss Light Source, CH-5232 Villigen PSI, Switzerland}
\author{T. Takahashi}
\affiliation{Department of Physics, Tohoku University, Sendai 980-8578, Japan}
\affiliation{WPI Research Center, Advanced Institute for Materials Research, Tohoku University, Sendai 980-8577, Japan}
\author{H. Ding}\email{dingh@iphy.ac.cn}
\affiliation{Beijing National Laboratory for Condensed Matter Physics, and Institute of Physics, Chinese Academy of Sciences, Beijing 100190, China}

\date{\today}

%\begin{minipage}[t]{6.8in}
\begin{abstract}
We used high-energy resolution angle-resolved photoemission spectroscopy to extract the momentum dependence of the superconducting gap of Ru-substituted Ba(Fe$_{0.75}$Ru$_{0.25}$)$_2$As$_2$ ($T_c = 15$ K). Despite a strong out-of-plane warping of the Fermi surface, the magnitude of the superconducting gap observed experimentally is nearly isotropic and independent of the out-of-plane momentum. More precisely, we respectively observed 5.7 meV and 4.5 meV superconducting gaps on the inner and outer $\Gamma$-centered hole Fermi surface pockets, whereas a 4.8 meV gap is recorded on the M-centered electron Fermi surface pockets. Our results are consistent with the $J_1-J_2$ model with a dominant antiferromagnetic exchange interaction between the next-nearest Fe neighbors. 
\end{abstract}


%\end{minipage}
\maketitle
%\narrowtext

The mechanism for Cooper pairing is the most important issue in high-$T_c$ superconductivity. Owing to its momentum resolution capabilities, angle-resolved photoemission spectroscopy (ARPES) is a very powerful tool to investigate the superconducting (SC) gap size and symmetry directly in the momentum space. Previous ARPES experiments revealed Fermi surface (FS) dependent and nearly isotropic SC gaps for both hole-doped \cite{Ding_EPL,L_Zhao} and electron-doped \cite{Terashima_PNAS2009} 122-ferropnictides. In contrast, a circular horizontal node at $k_z$ = $\pi$ was recently reported on a hole FS pocket with strong three-dimensional (3D) character in isovalent P-substituted BaFe$_2$(As$_{0.7}$P$_{0.3}$)$_2$ \cite{Y_Zhang_NaturePhys2012}. Whether such feature is unique or common to other isovalently-substituted BaFe$_2$As$_2$ materials is still under intense debate.  Ba(Fe$_{1-x}$Ru$_{x}$)$_2$As$_2$ is another isovalent substituted system with maximum critical temperatures $T_c$ around 20 K \cite{Thaler_PRB2010Ru, Albenque_prb2010,Eom_PRB85, J_Xing_Ru2012}. As with BaFe$_2$(As$_{0.7}$P$_{0.3}$)$_2$, previous ARPES studies of the electronic band structure indicate that this system exhibits a pronounced 3D character \cite{Brouet_PRL105,Dhaka_prl2012,Nan_XuPRB86}, thus raising the possibility of SC gap nodes. Indeed, recent measurements of a finite residual thermal conductivity suggest the presence of nodes. Unfortunately, samples of Ba(Fe$_{1-x}$Ru$_{x}$)$_2$As$_2$ usually show only a partial SC volume fraction detrimental to a determination of the momentum-resolved SC gap by ARPES.  

In this letter, we report high-energy resolution ARPES results on SC Ba(Fe$_{0.75}$Ru$_{0.25}$)$_2$As$_2$ (T$_c = 15$ K) recorded in the whole 3D Brillouin zone (BZ) by tuning the incident photon energy ($h\nu$). Due to improved sample quality, we observed the opening of a SC gap. We demonstrate that the SC gap size at this doping level is nearly isotropic on each FS, and slightly FS-dependent. Interestingly, the global gap structure in this material can be described by a single gap function derived from a strong coupling approach.

Large single crystals of Ba(Fe$_{0.75}$Ru$_{0.25}$)$_2$As$_2$ were grown by the self-flux method \cite{J_Xing_Ru2012}. ARPES measurements were performed at the 1-cubed ARPES end-station of BESSY and at Swiss Light Source beamline SIS using a VG-Scienta R4000 electron analyzer with $h\nu$ ranging from 22 to 64 eV, and at Tohoku University using a VG-Scienta SES2002 electron analyzer with a He discharge lamp ($h\nu$ = 21.218 eV). The angular resolution was set to 0.2$^\textrm{o}$ and the energy resolution to 4-7 meV for the SC gap measurements at $T = 0.9$ K in Bessy, and $T=6$ K in Tohuku University. Clean surfaces for the ARPES measurements were obtained by cleaving crystals \emph{in situ} in a working vacuum better than 5 $\times$ 10$^{-11}$ Torr. We label the momentum values with respect to the 1 Fe/unit cell BZ, and use $c'=c/2$ as the distance between two Fe planes.

\begin{figure}[!t]
\begin{center}
\includegraphics[width=3.4in]{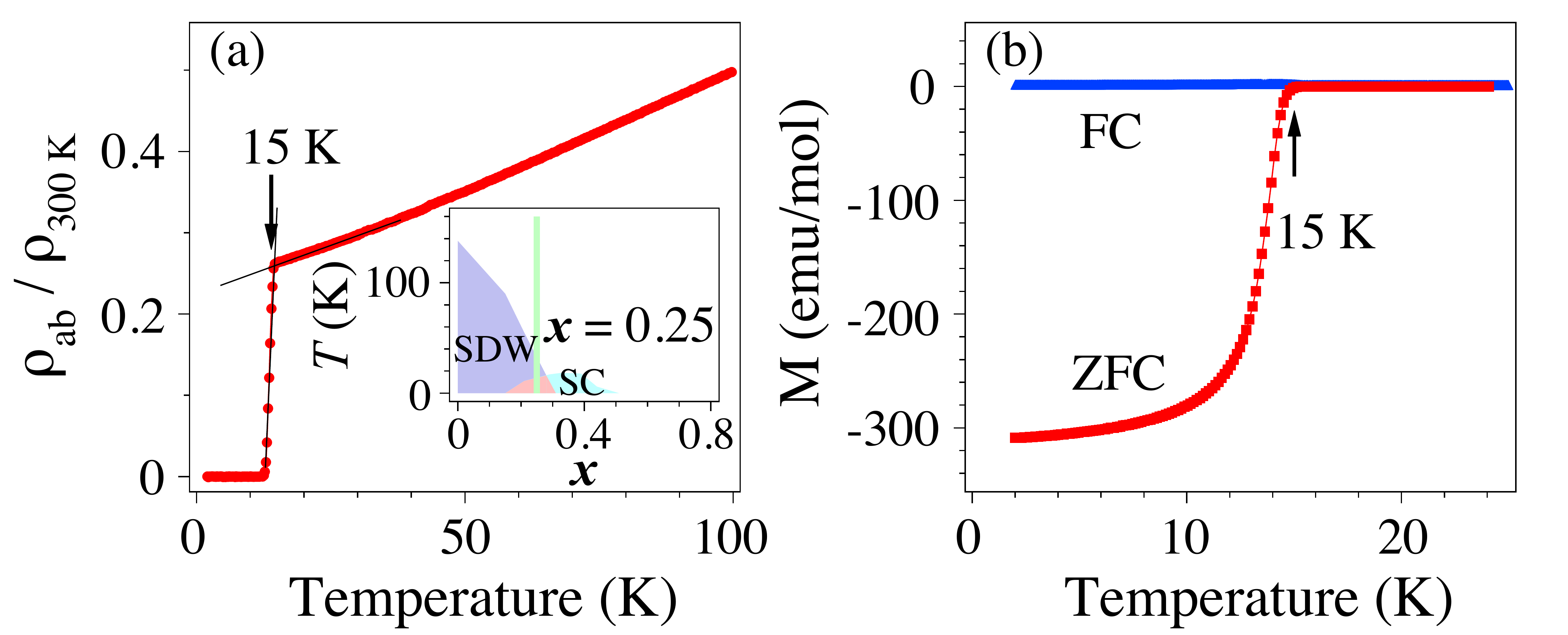}
\end{center}
\caption{\label{Fig1_transport}(Color online): (a) Temperature dependence of the normalized resistivity of  Ba(Fe$_{0.75}$Ru$_{0.25}$)$_2$As$_2$. Inset: phase diagram of Ba(Fe$_{1-x}$Ru$_{x}$)$_2$As$_2$. The green bar indicates the composition of the samples measured. (b) Temperature dependence of the DC magnetization of a Ba(Fe$_{0.75}$Ru$_{0.25}$)$_2$As$_2$ sample. A DC field of 10 Oe was applied in the measurements in the zero-field-cooling (ZFC) and field-cooling (FC) modes.}
\label{fig1_transport}
\end{figure}

Figures \ref{fig1_transport}(a) and \ref{fig1_transport}(b) show the temperature dependence of the resistivity and the DC magnetization of a Ba(Fe$_{0.75}$Ru$_{0.25}$)$_2$As$_2$ sample, respectively. Both the resistivity and zero-field cooled magnetization curves indicate an onset of superconductivity at $\sim 15$ K. Such $T_c$ value is consistent with the nominal composition of the sample, as indicated in the phase diagram shown in the inset of Fig. \ref{fig1_transport}(a). The resistivity goes down to zero within 2 K whereas the absolute value of the magnetization becomes large and tends to saturate at low temperature, suggesting that superconductivity is established in most of the bulk material. Yet, the saturation is not complete and the SC volume fraction is possibly slightly smaller than 100\%.

\begin{figure}[!t]
\begin{center}
\includegraphics[width=3.4in]{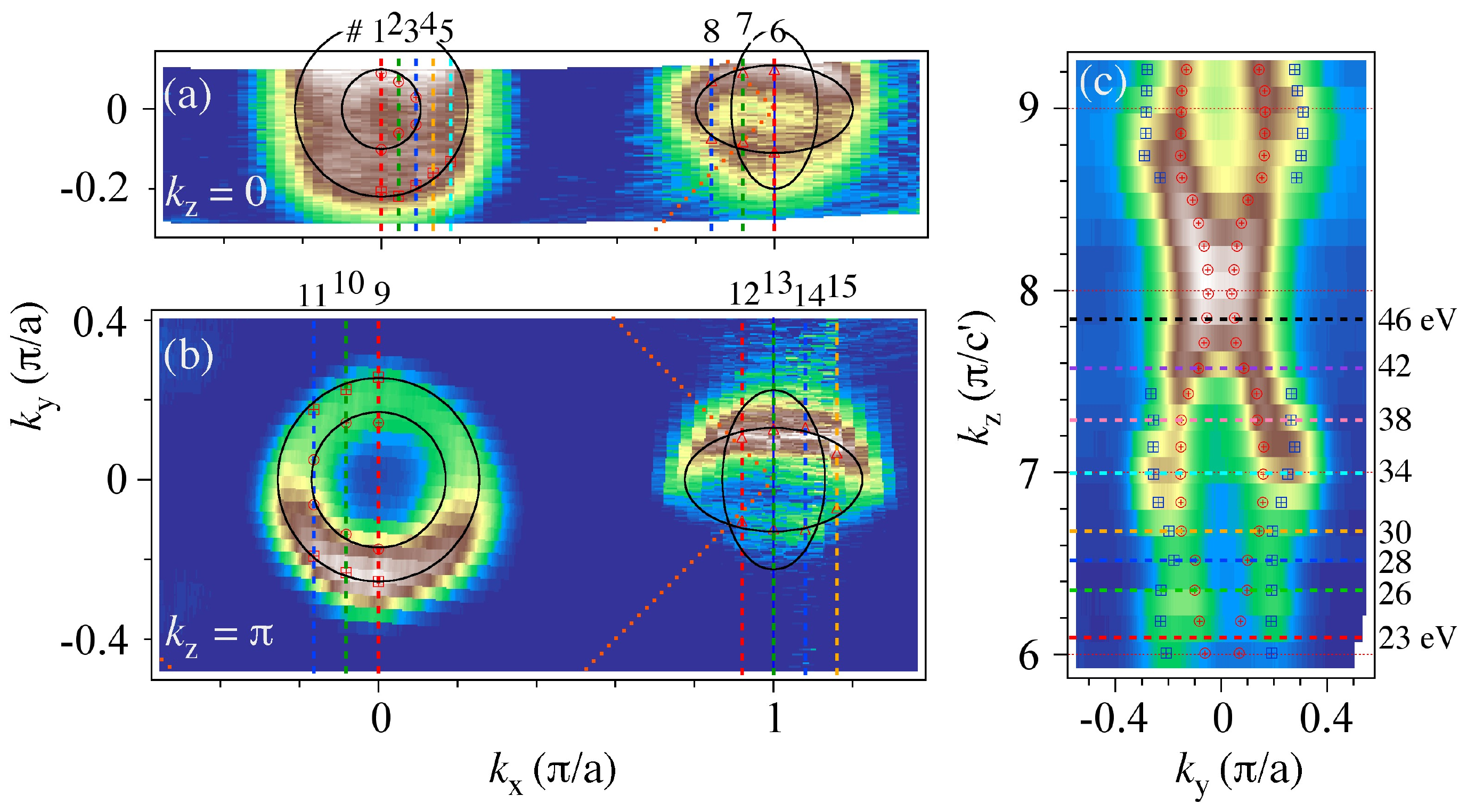}
\end{center}
\caption{\label{Fig2_FSs}(Color online) (a) and (b) FS mappings obtained by integrating the ARPES intensity within $E_F\pm10$ meV at (21.218 eV) $k_z=0$ and (34 eV) $k_z=\pi$, respectively. The black curves indicate the various FSs. (c) FS mapping in the $k_x$-$k_z$ plane obtained by tuning $h\nu$. The dashed lines and symbols indicate gap measurement cuts and $k_F$ locations.}
\label{fig2}
\end{figure}

Figures \ref{Fig2_FSs}(a) and \ref{Fig2_FSs}(b) show the FS mappings obtained by integrating the ARPES intensity within $E_F\pm10$ meV recorded with $h\nu=21.218$ eV and $h\nu=34$ eV, respectively. In agreement with our previous study on Ru-substituted BaFe$_2$As$_2$ \cite{Nan_XuPRB86}, we observed two almost circular hole FS pockets centered at the $\Gamma$ point, namely $\alpha$ and $\beta$, and two elliptical-like electron FS pockets centered at the M point, which hybridize to form the $\gamma$ and $\delta$ FS pockets. From the Fermi wave vector ($k_F$) positions determined experimentally by fitting the momentum distribution curves (MDCs) with Lorentz functions, which are appended in Figs. \ref{Fig2_FSs}(a) and \ref{Fig2_FSs}(b), we can trace the FS of this material. The contrast between the FS sizes obtained at 21.218 eV and 34 eV is important and suggests a strong out-of-plane momentum $k_z$ dependence. To confirm this effect, we performed ARPES experiments over a wide $h\nu$ range. The corresponding FS mapping of the $k_x$-$k_z$ plane is displayed in Fig. \ref{Fig2_FSs}(c), where we converted $h\nu$ into $k_z$ by using the nearly-free electron approximation \cite{DamascelliPScrypta2004} with an inner potential $V_0$ of 15 eV. As reported previously \cite{Brouet_PRL105,Dhaka_prl2012,Nan_XuPRB86}, the electronic structure of Ru-substituted BaFe$_2$As$_2$ compounds exhibits an enhanced 3D character as compared with K-doped and Co-doped samples that is mainly attributed the more specially extended nature of the Ru $4d$ orbitals. From Fig. \ref{Fig2_FSs}(c), we determine that $h\nu=21.218$ eV is very close to $k_z=0$ whereas 34 eV corresponds to $k_z=\pi/c^{\prime}$.  We also note that for this relatively low $Ru$ substitution level, the $\alpha$ band crosses the Fermi level ($E_F$) at all $k_z$ values.

\begin{figure*}[!t]
\begin{center}
\includegraphics[width=7in]{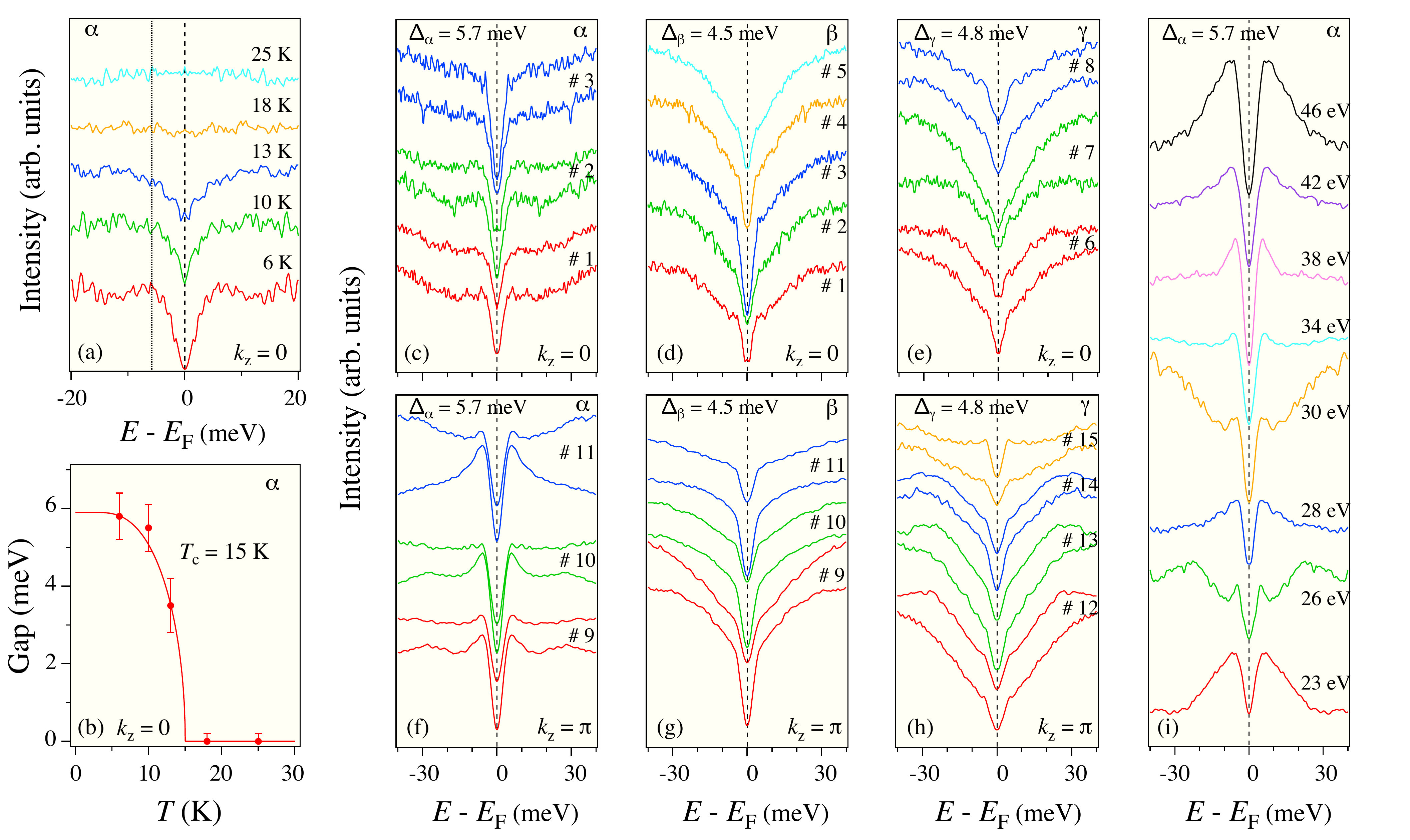}
\end{center}
\caption{\label{fig3_gap_structure}(Color online)  (a) Temperature dependence of the symmetrized EDC spectrum at a $k_F$ point of the $\alpha$ band. (b) Temperature dependence of the measured SC gap on the same band. The red curve corresponds to a fit using the BCS theory. (c)-(e) Symmetrized EDCs in the SC state (6 K) measured at various $k_F$ points in the $k_z = 0$ plane along the $\alpha$, $\beta$, and $\gamma/\delta$ FSs, respectively. The labels correspond to the cuts in Figs. \ref{Fig2_FSs}(a). (f)-(h) Same as (c)-(e) but measured at 0.9 K in the $k_z=\pi$ plane. (i) Symmetrized EDCs in the SC state (0.9 K) measured at various $k_F$ points along the $\alpha$ FS with various $h\nu$ values.}
\end{figure*}

We now turn our attention to the SC gap of Ba(Fe$_{0.75}$Ru$_{0.25}$)$_2$As$_2$. In order to detect the possible presence of nodes, we recorded high-energy resolution measurements along the various cuts indicated in Fig. \ref{Fig2_FSs}, which span the whole 3D BZ. The results are summarized in Fig. \ref{fig3_gap_structure}. Following a common practice, all the $k_F$ energy distribution curves (EDCs) have been symmetrized in order to approximately remove the Fermi function cutoff. In Fig. \ref{fig3_gap_structure}(a), we show the temperature evolution of the symmetrized EDC at one $k_F$ location on the $\alpha$ band. At 6 K, the symmetrized EDC spectrum exhibits a peak around 6 meV accompanied by a strong lost of intensity at $E_F$ that characterizes the opening of a SC gap. With temperature increasing, the peak broadens and the SC gap fills in. Both disappear between 13 K and 18 K, in agreement with the 15 K SC transition determined for our sample. We display in Fig. \ref{fig3_gap_structure}(b) the temperature dependence of the corresponding gap size.

The symmetrized EDCs recorded at 6 K along the $\alpha$, $\beta$ and $\gamma$/$\delta$ FSs at $k_z$ = 0 are shown in Figs. \ref{fig3_gap_structure} (c)-(e), respectively. As with most of the ARPES measurements on the 122 family of SC ferropnictides \cite{RichardRoPP2011}, the SC gap along each FS remains constant within experimental uncertainties. Hence, constant SC gaps of $\Delta_{\alpha}^{k_z=0}=5.7$ meV, $\Delta_{\beta}^{k_z=0}=4.5$ meV and $\Delta_{\gamma}^{k_z=0}=4.8$ meV are extracted along the $\alpha$, $\beta$ and $\gamma$ FSs, respectively. Similar results are obtained at 0.9 K for $k_z=\pi$, as shown in Figs. \ref{fig3_gap_structure} (f)-(h). Indeed, the same SC gap values are found experimentally, and thus $\Delta_{\alpha}^{k_z=\pi}=5.7$ meV, $\Delta_{\beta}^{k_z=\pi}=4.5$ meV and $\Delta_{\gamma}^{k_z=\pi}=4.8$ meV. These values lead respectively to $2\Delta$/$k_BT_c$ ratios of $9.0$, $7.1$ and $7.6$. Although experimental uncertainties would afford for small in-plane anisotropies of the SC gap, our results are inconsistent with the presence of vertical nodes.

\begin{figure}[!t]
\begin{center}
\includegraphics[width=3.4in]{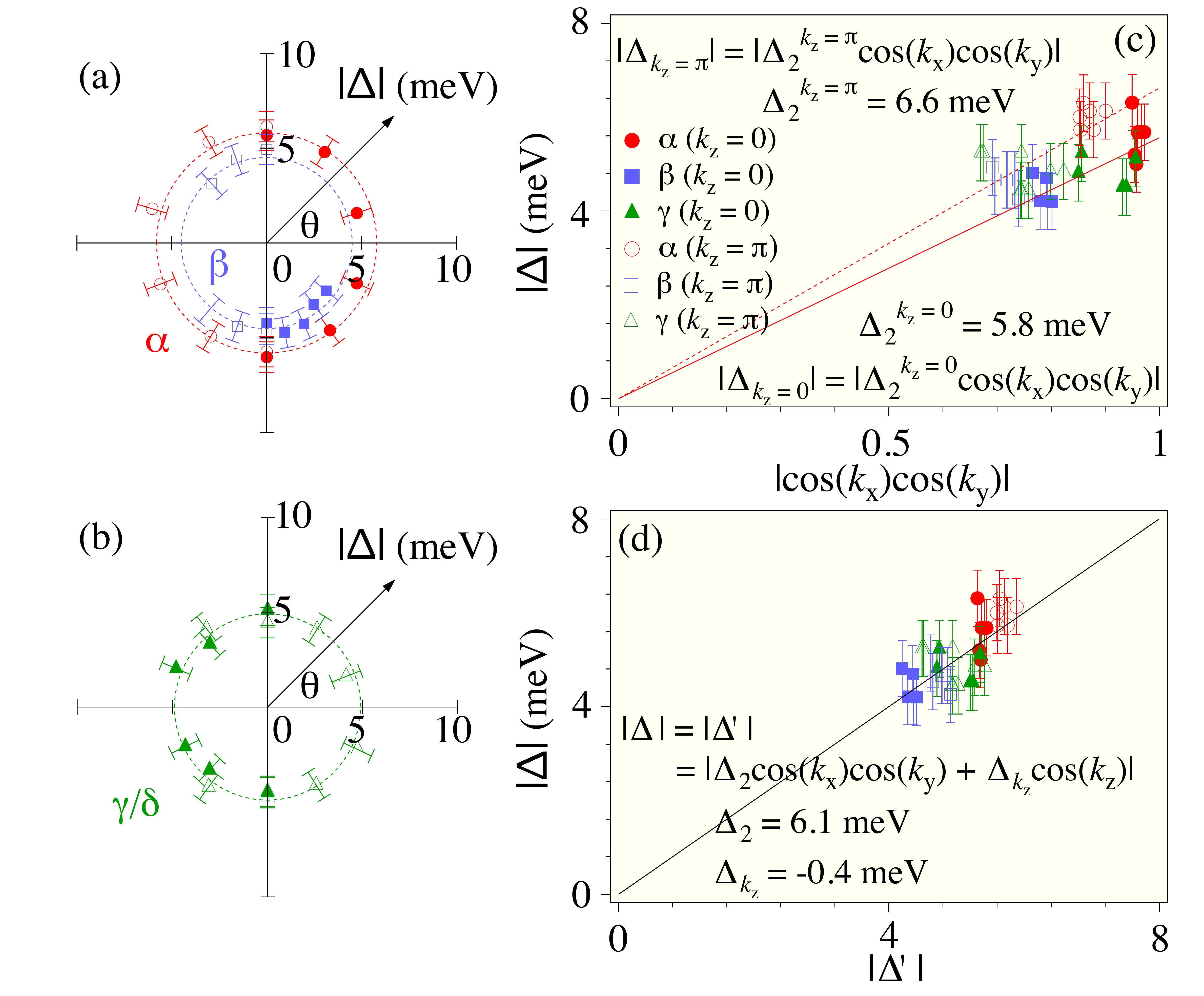}
\end{center}
\caption{\label{Fig4_DMFT}(Color online) (a)-(b) Polar plots of the SC gap size at 6 K for the $\alpha$, $\beta$ and $\gamma/\delta$ FSs as a function of $\theta$. Dashed circles show the averaged SC gap values on each FS. (c) SC gap size as a function of $|\cos k_x \cos k_y|$. The solid and dashed line represent fit for gap values at $k_z = 0$ and $k_z = \pi$, respectively. (d) SC gap size as a function of $|\Delta_2\cos k_x\cos k_y + \Delta_{k_z}\cos k_z|$ using simultaneously the data for both the $k_z=0$ and $k_z=\pi$ planes.}
\label{fig4_gap_summary}
\end{figure}

To check the possibility of horizontal nodes, we also performed high-energy resolution ARPES measurements along the warped $k_z$ FS of Ba(Fe$_{0.75}$Ru$_{0.25}$)$_2$As$_2$. The $h\nu$ dependence of the symmetrized EDCs at the $k_F$ positions associated with the $\alpha$ FS are plotted in  Fig. \ref{fig3_gap_structure}(i). Despite the large $k_z$ warping of the $\alpha$ band, our results indicate that the corresponding SC gap is practically constant at all measured $h\nu$ values, suggesting a 3D isotropic SC gap and ruling out the possibility of horizontal node for this band. In Figs. \ref{fig4_gap_summary}(a) and \ref{fig4_gap_summary}(b), we use the polar representation to illustrate the isotropicity of the SC gap observed for the $\Gamma$-centered and M-centered FS pockets, respectively. Even though our ARPES measurements do not allow us to make a conclusive statement on the possibility of horizontal node for the other bands, the equivalence of the SC gap in the $k_z=0$ and $k_z=\pi$ planes, which would be the likely possibilities for the occurrence of such node, strongly suggests that they do not exist.  Assuming that samples are not phase-separated, zero-field thermal conductivity is a very sensitive technique to detect the presence of nodes in the SC gap function, and the agreement with ARPES measurements is usually good \cite{RichardRoPP2011}. A recent thermal conductivity study suggests the presence of nodes in Ba(Fe$_{1-x}$Ru$_{x}$)$_2$As$_2$ \cite{X_Qiu_PRX2012}. Although the phase homogeneity of these materials is usually a matter of concern and that one cannot rule out phase separation, the current ARPES study would thus limit the latter observation to accidental nodes at momentum locations that remain undefined. 

The nearly isotropic SC gaps observed by ARPES for many Fe-based SC materials \cite{RichardRoPP2011} have serious implications for the underlying pairing mechanism \cite{Y_Huang_AIP2012}. In particular, they are inconsistent with the momentum-dependent quasi-nesting interactions promoted as the driving pairing force in weak coupling approaches \cite{MazinPhysicaC2009, Graser_NJP2009}. This is especially true for materials showing a large FS warping such as Ba(Fe$_{0.75}$Ru$_{0.25}$)$_2$As$_2$. Consequently, we rather analyze our results in the context of the strong coupling approach and the effective $J_1$-$J_2$ model \cite{SeoPRL2008, C_Fang_PRX2011, HuJP_SR2012, ZhouYi_EPL2011, HuJP_PRX2012}, as done previously for other Fe-based superconductors \cite{Nakayama_EPL2009,ZH_LiuPRB2011,Nakayama_PRB2011,YM_Xu_NPhys2011,XP_WangEPL2011,Y_Zhang_NaturePhys2012,H_Miao_PRB2012,UmezawaPRL2012}. Assuming that the $J_2$ parameter characterizing the antiferromagnetic exchange interaction between the next-nearest Fe neighbors dominates in the ferropnictides \cite{HuJP_SR2012}, we used a simple SC gap function of the form $|\Delta_2\cos k_x\cos k_y|$ to fit our data. In Fig. \ref{fig4_gap_summary}(c), we show the results when the $k_z=0$ and $k_z=\pi$ data are fit independently. Although both sets of data are well fit, they lead to slightly different SC gap parameters ($\Delta_2^{k_z = 0} = 5.8$ meV and $\Delta_2^{k_z = \pi} = 6.6$ meV). This discrepancy origins from the different size of the FSs in the two different $k_z$ planes. A better agreement is obtained when adding a gap parameter associated with inter-layer interactions, as done previously for Ba$_{0.6}$K$_{0.4}$Fe$_2$As$_2$ \cite{YM_Xu_NPhys2011}. The result of the simultaneous fit of the two sets of data to the SC gap function $|\Delta_2\cos k_x \cos k_y + \Delta_{k_z}\cos k_z|$ is displayed in Fig. \ref{fig4_gap_summary}(d). The fit leads to $\Delta_2 = 6.1$ meV and $\Delta_{k_z} = -0.4$ meV.

It is quite remarkable that the results obtained on the SC gap of Ba(Fe$_{0.75}$Ru$_{0.25}$)$_2$As$_2$ are not fundamentally different from those reported for its hole-doped and electron-doped SC cousins, which show much weaker FS warping along $k_z$. Furthermore, they reinforce the assumption that the horizontal node detected in BaFe$_2$(As$_{0.7}$P$_{0.3}$)$_2$, the other major family of isovalently-substituted 122-ferrorpnictide system, is not enforced by symmetry but rather accidental \cite{Y_Zhang_NaturePhys2012}. We caution that in our study the $k_F$ value associated to the $\alpha$ band is always smaller than that corresponding to the other $\Gamma$-centered hole bands. Our previous study indicates that this is no longer true for the high Ru-substitution side of the phase diagram \cite{Nan_XuPRB86}. This region could thus be more similar to the BaFe$_2$(As$_{0.7}$P$_{0.3}$)$_2$ system than the samples studied here. Yet, whether accidental horizontal nodes are also possible in this case remains an open question. 

In summary, we reported high-energy resolution ARPES results on isovalent Ru-substituted Ba(Fe$_{0.75}$Ru$_{0.25}$)$_2$As$_2$. We observed a nodeless and nearly isotropic SC gap that is slightly FS sheet dependent and remains constant at all $k_z$ values. This result indicates that this system is not fundamentally different from the related hole-doped and electron-doped 122 ferropnictide systems, thus pushing towards a rather universal picture for the Fe-based SC materials where the pairing interactions are essentially short-ranged.

\section*{ACKNOWLEDGMENTS}
We acknowledge J.-P. Hu for useful discussions, and Y. Tanaka, S. Souma and R. Ang for experimental assistance. This work was supported by grants from CAS (2010Y1JB6), MOST (2010CB923000 and 2011CBA001000, 2011CBA00102, 2012CB821403) and NSFC (10974175, 11004232 and 11034011/A0402) from China, JSPS, TRiP-JST, CREST-JST and MEXT of Japan, as well as the Cai Yuanpei program and the French ANR (project PNICTIDES). This work was supported by the Sino-Swiss Science and Technology Cooperation (project no. IZLCZ2 138954). This work is based in part upon research conducted at the Swiss Light Source, Paul Scherrer Institut, Villigen, Switzerland, and at BESSY, Helmholtz Zentrum, Berlin, Germany.

\bibliography{biblio_short}
%\bibliography{biblio_long}

\end{document}